\documentclass[12pt]{article}

\usepackage{amsmath,amssymb,amsfonts,amsthm,dsfont,dcolumn}
\usepackage[margin=-5pt]{caption}
\usepackage{graphicx,wrapfig}

\DeclareGraphicsExtensions{.pdf,.png,.jpg,.jpeg}

\newcolumntype{d}{D{.}{.}{0}}

\newlength{\textlength}
\newlength{\overlinelength}
\usepackage{color}

\newcounter{subequation}[equation]

\newcommand{\be}{\begin{equation}}
\newcommand{\ee}{\end{equation}}
\newcommand{\eel}[1]{\label{#1}\end{equation}}
\newcommand{\bea}{\begin{eqnarray}}
\newcommand{\eea}{\end{eqnarray}}
\newcommand{\eeal}[1]{\label{#1}\end{eqnarray}}
\makeatletter
\def\thesubequation{\theequation\@alph\c@subequation}
\def\@subeqnnum{{\rm (\thesubequation)}}
\def\slabel#1{\@bsphack\if@filesw {\let\thepage\relax
   \xdef\@gtempa{\write\@auxout{\string
      \newlabel{#1}{{\thesubequation}{\thepage}}}}}\@gtempa
   \if@nobreak \ifvmode\nobreak\fi\fi\fi\@esphack}
\def\subeqnarray{\stepcounter{equation}
\let\@currentlabel=\theequation\global\c@subequation\@ne
\global\@eqnswtrue \global\@eqcnt\z@\tabskip\@centering\let\\=\@subeqncr

$$\halign to \displaywidth\bgroup\@eqnsel\hskip\@centering
  $\displaystyle\tabskip\z@{##}$&\global\@eqcnt\@ne
  \hskip 2\arraycolsep \hfil${##}$\hfil
  &\global\@eqcnt\tw@ \hskip 2\arraycolsep
  $\displaystyle\tabskip\z@{##}$\hfil
   \tabskip\@centering&\llap{##}\tabskip\z@\cr}
\def\endsubeqnarray{\@@subeqncr\egroup
                     $$\global\@ignoretrue}
\def\@subeqncr{{\ifnum0=`}\fi\@ifstar{\global\@eqpen\@M
    \@ysubeqncr}{\global\@eqpen\interdisplaylinepenalty \@ysubeqncr}}
\def\@ysubeqncr{\@ifnextchar [{\@xsubeqncr}{\@xsubeqncr[\z@]}}
\def\@xsubeqncr[#1]{\ifnum0=`{\fi}\@@subeqncr
   \noalign{\penalty\@eqpen\vskip\jot\vskip #1\relax}}
\def\@@subeqncr{\let\@tempa\relax
    \ifcase\@eqcnt \def\@tempa{& & &}\or \def\@tempa{& &}
      \else \def\@tempa{&}\fi
     \@tempa \if@eqnsw\@subeqnnum\refstepcounter{subequation}\fi
     \global\@eqnswtrue\global\@eqcnt\z@\cr}
\let\@ssubeqncr=\@subeqncr
\@namedef{subeqnarray*}{\def\@subeqncr{\nonumber\@ssubeqncr}\subeqnarray}

\@namedef{endsubeqnarray*}{\global\advance\c@equation\m@ne
                           \nonumber\endsubeqnarray}

\makeatletter \@addtoreset{equation}{section} \makeatother
\renewcommand{\theequation}{\thesection.\arabic{equation}}
\catcode`\@=11

\newcount\hour
\newcount\minute
\newtoks\amorpm \hour=\time\divide\hour by 60\minute
=\time{\multiply\hour by 60 \global\advance\minute by-\hour}
\edef\standardtime{{\ifnum\hour<12 \global\amorpm={am}
        \else\global\amorpm={pm}\advance\hour by-12 \fi
        \ifnum\hour=0 \hour=12 \fi
        \number\hour:\ifnum\minute<10
        0\fi\number\minute\the\amorpm}}
\edef\militarytime{\number\hour:\ifnum\minute<10 0\fi\number\minute}

\def\draftlabel#1{{\@bsphack\if@filesw {\let\thepage\relax
   \xdef\@gtempa{\write\@auxout{\string
      \newlabel{#1}{{\@currentlabel}{\thepage}}}}}\@gtempa
   \if@nobreak \ifvmode\nobreak\fi\fi\fi\@esphack}
        \gdef\@eqnlabel{#1}}
\def\@eqnlabel{}
\def\@vacuum{}
\def\marginnote#1{}
\def\draftmarginnote#1{\marginpar{\raggedright\scriptsize\tt#1}}
\def\draft{
        \pagestyle{plain}
        \overfullrule=2pt
        \oddsidemargin -.5truein
        \def\@oddhead{\sl \phantom{\today\quad\militarytime} \hfil
        \smash{\Large\sl DRAFT} \hfil \today\quad\militarytime}
        \let\@evenhead\@oddhead
        \let\label=\draftlabel
        \let\marginnote=\draftmarginnote
        \def\ps@empty{\let\@mkboth\@gobbletwo
        \def\@oddfoot{\hfil \smash{\Large\sl DRAFT} \hfil}
        \let\@evenfoot\@oddhead}

\def\@eqnnum{(\theequation)\rlap{\kern\marginparsep\tt\@eqnlabel}
        \global\let\@eqnlabel\@vacuum}  }

\renewcommand{\theequation}{\thesection.\arabic{equation}}

\def\appendix#1{
  \addtocounter{section}{-3}
  \setcounter{equation}{0}
  \renewcommand{\thesection}{\Alph{section}}
  \section*{Appendix \thesection\protect\indent \parbox[t]{11.15cm}
  {#1} }
  \addcontentsline{toc}{section}{Appendix \thesection\ \ \ #1}
  }
\def\be{\begin{equation}}
\def\ee{\end{equation}}

\date{}
\begin{document}
%\draft

\begin{titlepage}

\begin{center}
%\vskip 2.5 cm \vskip 1 cm

{\Large \bf   A Quantum Rosetta Stone for the Information Paradox}

\vskip .7 cm

\vskip 1 cm

{\large  Leopoldo A. Pando Zayas}

\end{center}

\vskip .4cm \centerline{\it  Michigan Center for Theoretical
Physics}
\centerline{ \it Randall Laboratory of Physics, The University of
Michigan}
\centerline{\it Ann Arbor, MI 48109-1120}

\vskip .4cm
\centerline{ \it }
\centerline{\it  }

\vskip 1 cm

\vskip 1.5 cm

\begin{abstract}
The black hole information loss paradox epitomizes the contradictions between general relativity and quantum field theory. 
The AdS/CFT correspondence provides an implicit answer for the information loss paradox in black hole physics by equating a gravity theory with an explicitly unitary field theory. Gravitational collapse in asymptotically AdS spacetimes is generically turbulent. Given that the mechanism to read out the information about correlations functions in the field theory side is plagued by deterministic classical chaos, we argue that quantum chaos might provide the true Rosetta Stone for answering the information paradox in the context of the AdS/CFT correspondence.
\end{abstract}

\vskip 1 cm

\vskip 1.5 cm
{\large \bf Essay awarded Honorable Mention in the Gravity Research Foundation 2014 Awards for Essays on
Gravitation}

\end{titlepage}

\section{Introduction.}

The process of black hole formation and evaporation leads to many important conflicts in the interplay between quantum field theory and general relativity. No-hair theorems imply that most information about the collapsing body is lost from the outside region. The discovery that black holes radiate with a perfectly thermal featureless spectrum \cite{Hawking:1974sw} leads to the question of whether the information about the collapsing body is lost with the corresponding lost of unitarity -- a basic tenet of quantum theory --  or whether this information is somehow retrievable \cite{Hawking:1976ra}. This conundrum is known as the  black hole information loss paradox, it  best epitomizes the conflict between quantum field theory and general relativity  and has puzzled researchers for about forty years (see for example \cite{Preskill:1992tc} and more recently its articulation in the language of firewalls \cite{Almheiri:2012rt}).

The information loss paradox is assumed to be implicitly resolved in the context of the AdS/CFT correspondence \cite{Maldacena:1997re} where a gravity theory is equated to an explicitly unitary field theory.  The implicit resolution of the black hole information paradox in the context of the AdS/CFT \cite{Hawking:2005kf} still leaves us with the daunting question of  {\it how } exactly the paradox gets resolved and by what means information is retrieved from the black hole during its formation and evaporation.

Science has a way of teaching us through puzzles like the information loss paradox. We expect that the ultimate resolution will imply profound changes in our conceptual understanding of gravity as a quantum theory. This is perhaps the main  motivation to study this puzzle. There have been various suggestions stating that the problem might be not a  matter of principle but a matter of practicality.

In this very same forum -- Gravity Research  Foundation Essay Competition -- in his first prize award winning essay of 1997, Myers advocated that, at least in the context of string theory, the states forming a black hole are already mixed \cite{Myers:1997qi}. Another suggested resolution emphasizing practical limitations was also put forward by Balasubramanian, Marolf and Rozali who argued, in the first prize winning essay for  2006, that the problem resides in that Planck level resolution is needed to understand the structure of a black hole \cite{Balasubramanian:2006iw}.

Hawking has recently put forward the latest argument in favor of practical limitations for retrieving information from black holes \cite{Hawking:2014tga}. He argued that since the gravitational collapse to form an asymptotically AdS black hole will in general be turbulent  the dual CFT on the boundary of AdS will be chaotic implying, therefore, that information will  be effectively lost, although there would be no loss of unitarity. This situation is standard in deterministic chaos where even though the equations are deterministic there is a practical impossibility to reliably predict the state of the dynamical system after a certain asymptotically large time.

In this essay I argue that given the evidence in favor of turbulent collapse and for the high sensitivity to the initial conditions in the classical limit of the AdS/CFT, the ultimate resolution of the information loss paradox  necessarily requires quantum chaos. Quantum chaos has found its way to central aspects in physical phenomena such as Anderson localization, and the spectrum of many  systems including hadrons and Rydberg atoms. I argue that quantum chaos must play a central role in the understanding of the black hole information loss paradox. 

\section{Gravitational collapse is turbulent.}
An important property of asymptotically $AdS$ spacetimes, which is different from our intuition in asymptotically flat spacetimes, is the presence of a timelike boundary at spatial and null infinity where suitable boundary conditions need to be prescribed. Perturbations reaching this boundary can be reflected inward and are allowed to continue their evolution. 

To set up some notation, we consider a scalar field $\phi$, in a background defined by the following asymptotically AdS${}_{d+1}$ spacetime:

\be
ds^2=\frac{\mathit{l}^2}{\cos^2(x)}\bigg[-A e^{-2\delta} dt^2 + A^{-1}dx^2 +\sin^2(x) \,\,d\Omega_{d-1}\bigg],
\ee

\noindent where $d\Omega_{d-1}^2$ is the metric  of the round $(d-1)$-sphere and the metric functions $A,\delta$ depend on $(t,x)$. Here $t$ is the time coordinate and $0 \leq x \leq \pi/2$ covers the spatial domain with the boundaries at the origin ($x=0$) and the spatial infinity ($x=\pi/2$). The scalar field dynamics is best described in terms of auxiliary  variables $\Pi= A^{-1}e^{\delta} \dot{\varphi}$ and $\Phi =\varphi'$, where $\varphi(t,x)$ is the scalar field; prime and dot means derivatives with respect to $x$ and $t$, respectively.  The simplest exact solution of the equations of motion is the pure AdS spacetime characterized by $\Pi=\Phi=0$, $\delta=0$ and $A=1$.

Gravitational collapse, in this approximation, is an initial value problem. A typical configuration for the scalar field is given by a spherical shell with a Gaussian distribution: $\Phi(0,x)=0, \Pi(0,x)=\epsilon \exp\left(-\tan^2 x/\sigma^2\right)$, the parameters $\epsilon$ and $\sigma$ can be changed to explore the properties of collapse. It was originally reported in \cite{Bizon:2011gg} and later confirmed by various groups that AdS space is unstable under arbitrarily
small generic perturbations since a black hole can form after many reflections off the AdS boundary for arbitrarily small amplitudes, $\epsilon$. It was also conjectured that this instability is triggered by a resonant mode
mixing which gives rise to diffusion of energy from low to high frequencies. In \cite{deOliveira:2012dt}, the power spectrum was shown to be compatible with wave turbulence thus providing a numerical description of the turbulent cascade. There is recent evidence that the situation is more subtle as some configurations seem to be stable but {\it generically} gravitational collapse in AdS is turbulent.

%%%%%%%%%%%%%%%%%%%%%%%%%%%%%%%%%%%%%%%%%%%%%%%%%%%%%%%%%%%%%%%%%
\section{A Garbled entry in the AdS/CFT dictionary}
%%%%%%%%%%%%%%%%%%%%%%%%%%%%%%%%%%%%%%%%%%%%%%%%%%%%%%%%%%%%%%%
The master equation of the AdS/CFT correspondence is the identification of the partition functions of two theories: one containing gravity and one a field theory. This master formula allows to compute correlators in the CFT by evaluating the gravity action with a particular prescription for the boundary values of the fields:

\be
Z_{Gravity}[\phi(x,r)|_{\partial}=\phi_0(x)]=\langle \exp\left(\int d^4x \phi_0(x) {\cal O}(x)\right)\rangle_{CFT},
\ee
where $\phi(x,r)$ is a gravity bulk field whose boundary value, $\phi_0(x)$,  defines a source for the operator ${\cal O}$ in the conformal field theory \cite{Witten:1998qj,Gubser:1998bc}.

In its most widely used semiclassical limit, the AdS/CFT instructs us to look for classical solutions corresponding to the saddle points of  $Z_{Gravity}$. For fields of low conformal dimension the corresponding objects are classical fields satisfying the Klein-Gordon (KG) equation \cite{Witten:1998qj,Gubser:1998bc}. Operators with parametrically larger conformal dimensions are described in the bulk by classical strings or branes.

The AdS/CFT for gravitational collapse leads naturally, through its master equation, to considering the  KG equation for a scalar field in the collapsing background. We consider the field $\Psi$  as a probe, that is,
not including its back-reaction on the background and satisfying the KG equation:

\begin{eqnarray}
\label{Eq:probe}
&-&e^{\delta}\cos^2\left(\frac{x}{\ell}\right)\partial_t \left(e^{\delta}A^{-1}\partial_t \Psi\right)+m^2  \Psi \nonumber \\
&+&e^\delta \frac{\cos(\frac{x}{\ell})}{\sin^2(\frac{x}{\ell})}\partial_x \left(e^{-\delta} A\sin^2(\frac{x}{\ell})\cos(\frac{x}{\ell}) \partial_x \Psi\right) =0,
\end{eqnarray}
where $A$ and $\delta$ determine the collapsing background and are read off from simulations discussed extensively in  \cite{deOliveira:2012dt}.

\begin{figure}[htp]
\begin{center}
\includegraphics[width=3.5in]{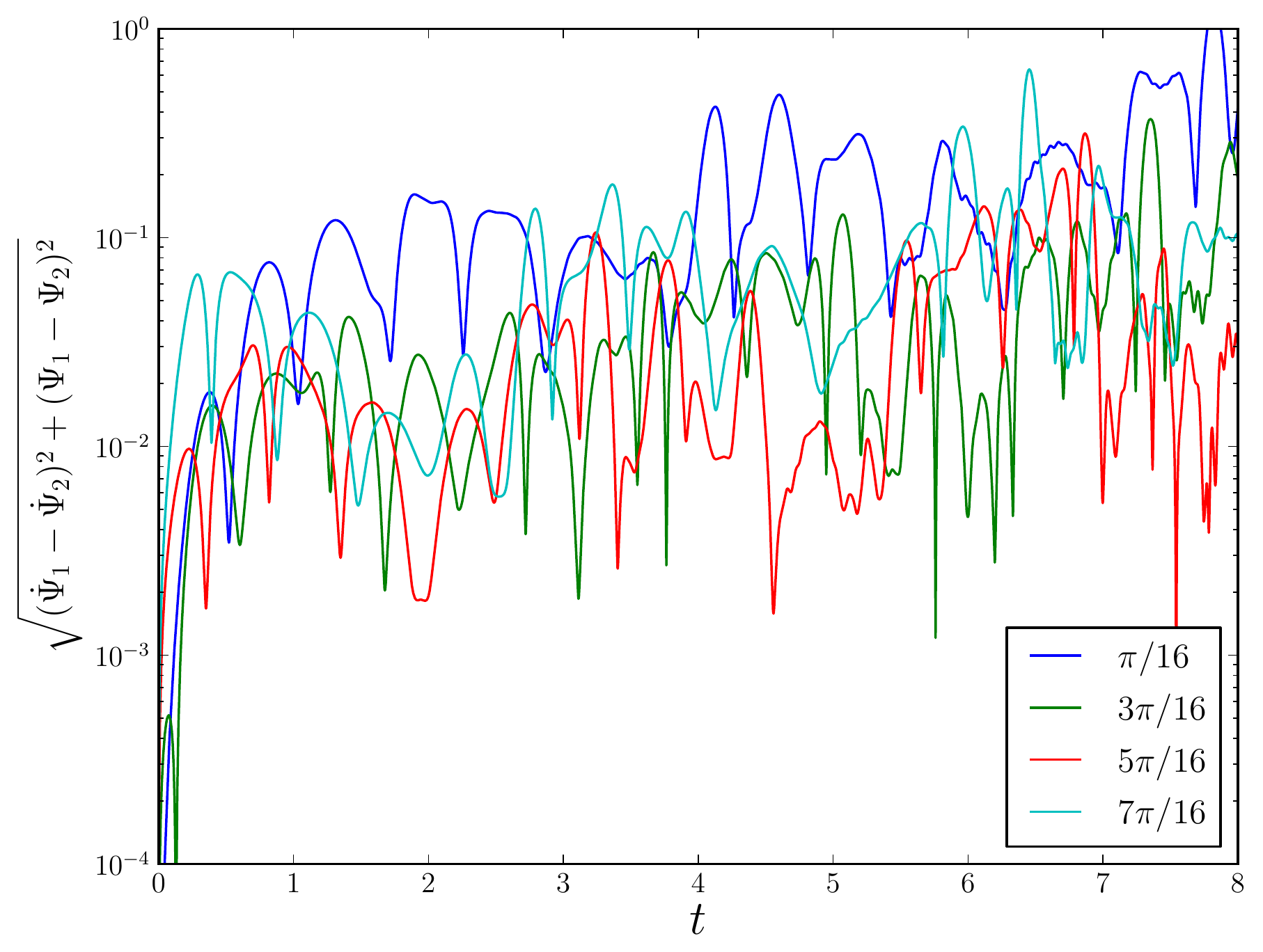}
\caption{\label{LE} For two collapsing simulations with amplitudes $\epsilon=0.4, 0.41$ we plot the phase space distance in logarithmic scale as a function of time. For all points the growth trend implies, according to our definition,  a positive Lyapunov exponent.}
\end{center}
\end{figure}

Although the AdS/CFT dictionary in the context of arbitrarily time-dependent configurations has not been rigorously formulated, the natural working assumption  is that sources and responses in the field theory are read from the asymptotic behavior
of the field $\Psi$. Spontaneous symmetry breaking is implemented through a boundary condition with no source. A
source is difficult to implement numerically because it corresponds to a non-normalizable mode. For numerical expediency and given
the hyperbolic nature of the
equation versus its more generic elliptic nature in time-independent situations of the original AdS/CFT prescription, we choose
to evolve and initial profile of the probe scalar field of the form: $\Psi(t=0, x)=\sin^3 (2x),\,\, \dot{\Psi}(t=0, x)=0 $. In \cite{Farahi:2014lta} various indicators for classical chaos were presented. In particular the phase space trajectories of the probe scalar field $(\Psi, \dot{\Psi})$ were presented. As further evidence in favor of sensitivity to the initial conditions the largest Lyapunov exponent was considered in \cite{Farahi:2014lta}. Modulo subtleties related to the situation corresponding to PDE, rather than to a dynamical system and problems going to asymptotically large times due to the finiteness of collapse time, a natural definition of Lyapunov exponent was used. Given a point in phase space, one examines  the following
quantity {\small $\lambda (T) = \ln \sqrt{(\Psi_{\epsilon_1}(x_0, T)-\Psi_{\epsilon_2}(x_0, T))^2
+ (\dot{\Psi}_{\epsilon_1}(x_0, T)-\dot{\Psi}_{\epsilon_2}(x_0, T))^2}$}. The largest Lyapunov exponent is defined as the
slope of the $(\lambda(T),T)$ graph.  With all the previous caveats mentioned,  the value of this quantity was explored for various points $x_0$'s and it was found to be positive, pointing to exponential sensitivity to the initial conditions. The results presented in figure \ref{LE}. Note that we have considered two nearby amplitudes $\epsilon_{1,2}=0.4, 0.41$.

Note that the meaning of a positive Lyapunov exponent in this problem is {\it sui generis}. Equation \ref{Eq:probe} is linear in the field $\Psi$, we not discussing sensitivity to the initial conditions in the evolution of $\Psi$. We have studied the response of $\Psi$ to a slight
change in the initial conditions that trigger gravitational collapse by considering collapse of the Einstein-scalar in the previous section with an initial Gaussian profile with amplitudes $\epsilon_1=0.40$ and $\epsilon_=0.41$. For these two nearby backgrounds we study the probe scalar equation $\Psi$. This protocol is  equivalent to asking, in the field theory side, whether
correlation functions extracted during a thermalization process remain
predictably close so as to permit full reconstruction even after accounting for a slight uncertainty in the amount of initially injected energy.

%%%%%%%%%%%%%%%%%%%%%%%%%%%%%%%%%%%%%%%%%%%%%%%%%%%%%%%%%%%%%%%%%%%%%%%%%%%%%%%%5
\section{ Towards a quantum Rosetta Stone for the Information Paradox}

What is now known as quantum chaos, that is, the study of properties of quantum systems whose classical limit is chaotic, was set in motion about a century ago when Einstein questioned the status of Born quantization $\int pdx =n\hbar$ in the context of classical systems that do not admit integrals of motion. There are by now various monographs on quantum chaos \cite{gutzwiller-quantum-chaos}.  As opposed to classical chaos, quantum chaos is no longer concerned with solutions of the classical equations of motion, their phase space properties and sensitivity to changes in initial conditions since such sensitivity does not exist in the quantum case. One of the main results pertains to the local statistics of the energy spectrum. In particular, to the level spacing
distribution $P(s)$, which is the distribution function of nearest-neighbor spacings $\lambda_{n+1}-\lambda_n$ as
we run over all levels. If the classical dynamics is integrable, then $P(s)$ coincides with the corresponding quantity for a sequence
of uncorrelated levels (Poisson ensemble) with the same mean spacing: $P(s)=c e^{-cs}$. If the classical dynamics is chaotic, then $P(s)$ coincides with the corresponding quantity for the eigenvalues of a suitable ensemble of random matrices. This universality at the quantum level is truly humbling.

%%%%%%%%%%%%%%%%%%%%%%%%%%%%%%%%%%%%%%%%%%%%%%%%%%%%%%%%%%%%%%%%%%%%%%%%%%%%%%%%%%%%%%%%%%
%\subsection{The kicked quantum rotor}
%%%%%%%%%%%%%%%%%%%%%%%%%%%%%%%%%%%%%%%%%%%%%%%%%%%%%%%%%%%%%%%%%%%%%%%%%%%%%%%%%%%%%%%%%

A particularly telling example from quantum chaos is the kicked rotor whose Hamiltonian is given by:
\be
H(x,p,t)=\frac{1}{2}p^2 +K\cos(x)\sum\limits_{n=0}^\infty \delta(t-n),
\ee
where $ \delta$  is the Dirac delta function, $x$  is the angular position, $ p$  is the momentum, and $\textstyle K$  is the kicking strength. This system displays diffusive energy growth (linear in time) 

\be
E_n=\frac{K^2}{4}n.
\ee

What might be particularly relevant to black hole physics is that classical chaos is suppressed in quantum systems as  orginally discovered numerically Casatti, Chirikov, Izraelev and Ford for this very system \cite{Chirikov}. Quantum mechanically the kicked rotor gains energy as in the classical case only for a short time, after which the diffusion is suppressed. This phenomenon has found applications in many areas including mapping the kicked rotor onto the Anderson localization problem, which considers electron transport in a disordered crystal \cite{FGP}.

Some aspects of quantum chaos have been recently considered in the context of string theory \cite{PandoZayas:2012ig}. In string theory the Virasoro constraint provides the analog to the Schr\"odinger equation.  It leads to the mass shell condition and is usually written as $\left({\cal H}=(L_0-a)\right)|\Psi>=0$, where $L_0$ is a Virasoro generator and $a$ is a constant resulting from normal ordering.  In this framework we think of $|\Psi>$ as a the precise analog of the wave function in quantum mechanics. To make concrete progress further approximations are usually required. A natural one is the so called  mini-superspace, an approximation where only modes corresponding to the center of mass are retained. The minisuperspace idea was originally introduced in the context of quantum cosmology \cite{Hartle:1983ai}. Note that equation (\ref{Eq:probe}) would correspond to allowing only the center of mass degrees of freedom. The work presented in \cite{PandoZayas:2012ig} and \cite{Basu:2013uva} kept some degrees of freedom related to the spatial extension of the string and was able to produce a spectrum for holographic hadrons that matches qualitatively those of real hadrons following the Particle Data Table.

%%%%%%%%%%%%%%%%%%%%%%%%%%%%%%%%%%%%%%%%%%%%%%%%%%%%%%%%%%%%%%%%%%%%%%%%%%%%%%%%%%%%
%\subsection{Chaotic Quantum Field Theory}
%%%%%%%%%%%%%%%%%%%%%%%%%%%%%%%%%%%%%%%%%%%%%%%%%%%%%%%%%%%%%%%%

{\bf Towards Chaotic Quantum Field Theory in Curved Spaces} -- One natural question that the current state of affairs suggests is to reconsider Hawking's calculation \cite{Hawking:1974sw} in the context of chaotic classical dynamics. Classical solutions play an important role in the Hawking process. The paradigm of using  Bogolyubov transformations to relate modes at different regions of spacetime relies critically in the form of the solutions in various regions; these solutions are now generically chaotic, as indicated in this essay.

{\it We should take the lessons that quantum chaos has already taught us, extend them to the context of quantum field theory in curved spacetimes and use them  in conjunction with the AdS/CFT correspondence as the Rosetta Stone that would allow us to decipher the puzzle of black hole information loss paradox.}

%%%%%%%%%%%%%%%%%%%%%%%%%%%%%%%%%%%%%%%%%%%%%%%%%%%%%%%%%%%%%%%%%%%%%%%%%%%%%%%%%%%%%%%%%%%%%
\section*{Acknowledgments}
%%%%%%%%%%%%%%%%%%%%%%%%%%%%%%%%%%%%%%%%%%%%%%%%%%%%%%%%%%%%%%%%%%%%%%%%%%%%%%%%%%%%%%
I thank my collaborators P. Basu, A. Farahi, D. Garfinkle, H. de Oliveira, D. Reichmann, E. Rodrigues and C. Terrero-Escalante. This work is  partially supported by Department of Energy under grant DE-FG02-95ER40899 to the University of Michigan.

\end{document}